\documentstyle[12pt]{article}
\newcommand{\bea}{\begin{eqnarray}}
\newcommand{\eea}{\end{eqnarray}}
\newcommand{\be}{\begin{equation}}
\newcommand{\ee}{\end{equation}}

\newcommand{\rf}[1]{(\ref{#1})}

\begin{document}

\begin{center}
{\bf EXISTENCE AND PROPERTIES OF THE $f_0(665)$ STATE AND CHIRAL SYMMETRY}
\footnote{
This work was supported by the Grant Program of Plenipotentiary of Slovak
Republic at JINR.  Yu.S. and M.N. were supported in part by the Slovak
Scientific Grant Agency, Grant VEGA No. 2/7175/20; and D.K., by Grant VEGA
No. 2/5085/99.}

\bigskip

\underline{Yu.S. Surovtsev}$^a$, D.~Krupa$^b$, M.~Nagy$^b$\\
\end{center}
$~~~~~~~^a$ {\it Bogoliubov Laboratory of Theoretical Physics, JINR, Dubna
141 980, Russia}\\
$~~~~~~~~^b$ {\it Institute of Physics, SAS, D\'ubravsk\'a cesta 9, 842 28
Bratislava, Slovakia}
\begin{abstract}
\baselineskip 10pt
On the basis of a simultaneous description of the isoscalar $s$-wave
of $\pi\pi$ scattering (from the threshold up to 1.9 GeV) and of
$\pi\pi\to K\overline{K}$ process (from the threshold to $\sim$ 1.4 GeV) in
the model-independent approach, it is shown that there exists the $f_0(665)$
state with properies of the $\sigma$-meson, the glueball nature of
$f_0(1500)$ is indicated, and the $f_0(1370)$ is assigned mainly to
$s{\bar s}$ state. The coupling constants of the observed states with
$\pi\pi$ and $K\overline{K}$ systems and scattering lengths $a_0^0(\pi\pi)$
and $a_0^0(K\overline{K})$ are calculated. The existence of the $f_0(665)$
state and the obtained $\pi\pi$-scattering length
($a_0^0\approx 0.27 m_{\pi^+}^{-1}$) seem to suggest the linear realization
of chiral symmetry.
\\

{\bf Key-words:} analyticity, unitarity, uniformization, multichannel
resonance, scalar meson, QCD nature
\end{abstract}

\noindent{\bf 1.}
The scalar-meson sector causes a lot of questions up to now. A strong model
dependence of information on these objects (as wide multichannel states)
obtained from experimental data makes sometimes the drawn conclusions
doubtful.
{\it E.g.}, we have shown \cite{KMS-nc96} that an inadequate description of
multichannel states gives not only their distorted parameters when analyzing
data but also can cause the fictitious states when one neglects important
channels. In this report, we show that a large background earlier obtained in
various analyses of the $s$-wave $\pi\pi$ scattering \cite{PDG-00} hides, in
reality, the $\sigma$-meson and the influence of the left-hand branch-point.
The state with $\sigma$-meson properties is required by most models (like the linear $\sigma$-models and
the Nambu -- Jona-Lasinio models \cite{NJL}) for spontaneous breaking of
chiral symmetry.

A model-independent information on multichannel states can be obtained only
on the basis of the first principles (analyticity and unitarity) immediately
applied to analyzing experimental data and using the mathematical fact that
the local behaviour of analytic functions determined on the Riemann surface
is governed by the nearest singularities on all sheets. Earlier, we have
proposed that method for 2- and 3-channel resonances and developed the
concept of standard clusters (poles on the Riemann surface) as a qualitative
characteristic of a state and a sufficient condition of its existence
\cite{KMS-nc96}. We outline this below for the 2-channel case. Then
we analyze simultaneously experimental data on the processes
$\pi\pi\to\pi\pi,K\overline{K}$ in the channel with $I^GJ^{PC}=0^+0^{++}$.
On the basis of obtained pole clusters for resonances, their coupling
constants with the considered channels and scattering lengths compared with
the results of various models, we conclude on the nature of the observed
states and on the mechanism of chiral-symmetry breaking.

\noindent{\bf 2. Two-Coupled-Channel Formalism.}
We consider the coupled processes of $\pi\pi$ and $K\overline{K}$
scattering and $\pi\pi\to K\overline{K}$. Therefore, we have the 2-channel
$S$-matrix determined on the 4-sheeted Riemann surface. The matrix
elements $S_{\alpha\beta}$, where $\alpha,\beta=1(\pi\pi), 2(K\overline{K})$,
have the right-hand (unitary) cuts along the real axis of the $s$-variable
complex plane ($s$ is the invariant total energy squared), starting at
$4m_\pi^2$ and $4m_K^2$, and the left-hand cuts, beginning at $s=0$ for
$S_{11}$ and at $4(m_K^2-m_\pi^2)$ for $S_{22}$ and $S_{12}$. The
Riemann-surface sheets are numbered according to the signs of analytic
continuations of the channel momenta ~$k_1=(s/4-m_\pi^2)^{1/2}$,
~$k_2=(s/4-m_K^2)^{1/2}$~ as follows:
signs $({\mbox{Im}}k_1,{\mbox{Im}}k_2)=++,-+,--,+-$ correspond to the sheets
I, II, III, IV.

The resonance representations on the Riemann surface are obtained with the
help of the formulae, expressing analytic continuations of the matrix
elements to unphysical sheets $S_{\alpha\beta}^L$ ($L=II,III,IV$) in
terms of those on the physical sheet $S_{\alpha\beta}^I$ which have only
zeros (beyond the real axis), corresponding to resonances:
\begin{eqnarray} \label{S_L}
&&S_{11}^{II}=\frac{1}{S_{11}^I},\qquad ~~~~S_{11}^{III}=\frac{S_{22}^I
}{\det S^I},
\qquad S_{11}^{IV}=\frac{\det S^I}{S_{22}^I},\nonumber\\
&&S_{22}^{II}=\frac{\det S^I}{S_{11}^I},\qquad S_{22}^{III}=\frac{S_{11
}^I}
{\det S^I},\qquad S_{22}^{IV}=\frac{1}{S_{22}^I},\\
&&S_{12}^{II}=\frac{iS_{12}^I}{S_{11}^I},\qquad ~~~S_{12}^{III}=\frac{-
S_{12}^I}
{\det S^I},\qquad S_{12}^{IV}=\frac{iS_{12}^I}{S_{22}^I},\nonumber
\end{eqnarray}
Here $\det S^I=S_{11}^I S_{22}^I-(S_{12}^I)^2$.
These formulae immediately give the resonance representation by poles and
zeros on the 4-sheeted Riemann surface. One must discriminate between three
types of 2-channel resonances described by a pair of conjugate zeros on sheet
I: ({\bf a}) in $S_{11}$, ({\bf b}) in $S_{22}$, ({\bf c}) in each of $S_{11}$
and $S_{22}$.
As seen from \rf{S_L}, to the resonances of types ({\bf a}) and ({\bf b})
one has to make correspond a pair of complex conjugate poles on sheet III,
shifted relative to a pair of poles on sheet II and IV, respectively.
To the states of type ({\bf c}) one must consider corresponding two pairs of
conjugate poles on sheet III.
A resonance of every type is represented by a pair of complex-conjugate
clusters (of poles and zeros on the Riemann surface) of size typical of
strong interactions.
The cluster kind is related to the state nature. The resonance, coupled
relatively more strongly to the $\pi\pi$ channel than to the
$K\overline{K}$ one, is described by the cluster of type ({\bf a}); in the
opposite case it is represented by the cluster of type ({\bf b}) (say, the
state with the dominant $s{\bar s}$ component); the flavour singlet
({\it e.g.} glueball) must be represented by the cluster of type ({\bf c}) as
a necessary condition.

For the simultaneous analysis of experimental data on the coupled processes
it is convenient to use the Le Couteur-Newton relations \cite{LN} expressing
the $S$-matrix elements of all coupled processes in terms of the Jost matrix
determinant $d(k_1,k_2)$, the real analytic function with the only
square-root branch-points at $k_i=0$.
To take into account these right-hand branch-points at $4m_\pi^2$ and
$4m_K^2$ and also the left-hand one at $s=0$, we use the uniformizing
variable
\begin{equation} \label{v}
v=\frac{m_K\sqrt{s-4m_\pi^2}+m_\pi\sqrt{s-4m_K^2}}{\sqrt{s(m_K^2-m_\pi^2)}}
\end{equation}
which maps the 4-sheeted Riemann surface onto the $v$-plane, divided into two
parts by a unit circle centered at the origin. Sheets I (II), III (IV)
are mapped onto the exterior (interior) of the unit disk on the upper and
lower $v$-half-plane, respectively. The physical region extends from point
$i$ on the imaginary axis ($\pi\pi$ threshold) along the unit circle clockwise
in the 1st quadrant to point 1 on the real axis ($K\overline{K}$ threshold)
and then along the real axis to point $b=\sqrt{(m_K+m_\pi)/(m_K-m_\pi)}$ into
which $s=\infty$ is mapped on the $v$-plane. The intervals
$(-\infty,-b],[-b^{-1},b^{-1}],[b,\infty)$ on the real axis are the images of
the corresponding edges of the left-hand cut of the $\pi\pi$-scattering
amplitude. The type ({\bf a}) resonance is represented in $S_{11}$ by two
pairs of the poles on the images of sheets II and III, symmetric to each
other with respect to the imaginary axis, and by zeros, symmetric to these
poles with respect to the unit circle.
On $v$-plane, the Le Couteur-Newton relations are
\begin{equation} \label{v:C-Newton}
S_{11}=\frac{d(-v^{-1})}{d(v)},\quad S_{22}=\frac{d(v^{-1})}{d(v)},
\quad S_{11}S_{22}-S_{12}^2=\frac{d(-v)}{d(v)}.
\end{equation}
The condition of the real analyticity implies $ ~d(-v^*)=d^* (v)~$
for all $v$, and the unitarity needs the following relations to hold true for
the physical $v$-values: $ ~|d(-v^{-1})|\leq |d(v)|$, $~~ |d(v^{-1})|\leq
|d(v)|,~$  $~|d(-v)|=|d(v)|$.

The $d(v)$-function that already does not possess branch-points is taken as
$~d=d_B d_{res}$ where $d_B$ is the part of the $K\overline{K}$ background
that does not contribute to the $\pi\pi$-scattering amplitude.
On the $v$-plane, $S_{11}$ has no cuts, however, the amplitudes of
$K\overline{K}\to\pi\pi,K\overline{K}$ processes do have the cuts on the
$v$-plane, which arise from the left-hand cut on the $s$-plane, starting at
the point $s=4(m_K^2-m_\pi^2)$. This left-hand cut will be neglected in the
Riemann-surface structure, and the contribution on the cut will be taken into
account in the $K\overline{K}$ background as a pole on the real $s$-axis on
sheet I in the sub-$K\overline{K}$-threshold region; on the $v$-plane, this
pole gives two poles on the unit circle in the upper half-plane, symmetric to
each other with respect to the imaginary axis, and two zeros, symmetric to
the poles with respect to the real axis. Therefore,
\begin{equation} \label{B_K}
d_B=v^{-4}(1-pv)^4(1+p^*v)^4,
\end{equation}
where $p$ is the position of zero on the unit circle.
The fourth power in (\ref{B_K}) follows from eqs.(\ref{S_L}) and from that,
for the $s$-channel process $\pi\pi\to K\overline{K}$, the crossing $u$- and
$t$-channels are the $\pi-K$ and $\overline{\pi}-K$ scattering (exchanges in
these channels contribute on the left-hand cut).
The function $d_{res}(v)$ represents the contribution of resonances and,
except for the point $v=0$, consists of the zeros of clusters:
\begin{equation} \label{d_res}
d_{res} =v^{-M}\prod_{n=1}^{M} (1-v_n^* v)(1+v_n v),
\end{equation}
where $M$ is the number of pairs of the conjugate zeros.

\noindent{\bf 3. Analysis of experimental data.}
We analyze simultaneously the available experimental data on the
$\pi\pi$-scattering \cite{Hyams} and process $\pi\pi\to K\overline{K}$
\cite{Wickl} in the channel with $I^GJ^{PC}=0^+0^{++}$. As the data, we use
the results of phase analyses which are given for phase shifts of the
amplitudes ($\delta_1$ and $\delta_{12}$) and for moduli of the $S$-matrix
elements $\eta_a=|S_{aa}|$ (a=1$\pi\pi$,2$K\overline{K}$) and $\xi=|S_{12}|$.
The 2-channel unitarity condition gives
~~$\eta_1=\eta_2=\eta,~~ \xi=(1-\eta^2)^{1/2},~~ \delta_{12}=
\delta_1+\delta_2$.

For a satisfactory description of the $s$-wave $\pi\pi$ scattering from the
threshold to 1.89 GeV and process $\pi\pi\to K\overline{K}$ from the threshold 
to $\sim$ 1.4 GeV, three resonances turned out to be sufficient (solution 1): 
the two ones of type ({\bf a}) ($f_0 (665)$ and $f_0 (980)$) and $f_0 (1500)$ 
of type ({\bf c}). 
{\it I.e.}, a minimum scenario of the simultaneous description of these two
processes does not require the ${f_0}(1370)$ resonance, therefore, if this 
meson exists, it must be relatively more weakly coupled to the $\pi\pi$ 
channel than to the $K\overline{K}$ one, {\it i.e.}, be described by the 
cluster of type ({\bf b}) (this would testify to the dominant $s{\bar s}$
component in this state). To confirm quantitatively this qualitative
conclusion \cite{skn-app}, we consider also the 2nd solution including the
${f_0}(1370)$ of type ({\bf b}). 

{\bf Solution 1:} For $\delta_1$ and $\eta$, 113 and 50 experimental points
\cite{Hyams}, respectively, are used; when rejecting the points at 0.61,
0.65, and 0.73 GeV for $\delta_1$ and at 0.99, 1.65, and 1.85 GeV for $\eta$
which give an anomalously large contribution to $\chi^2$, we obtain for
$\chi^2/\mbox{ndf}$ the values 2.7 and 0.72, respectively; the total
$\chi^2/\mbox{ndf}$ in the case of $\pi\pi$ scattering is 1.96.

The satisfactory description for $|S_{12}|$ is given from the threshold to
$\sim$ 1.4 GeV. Here 35 experimental points \cite{Wickl} are used;
$\chi^2/\mbox{ndf}\approx 1.11$ when eliminating the points at 1.002, 1.265,
and 1.287 GeV (with an especially large contribution to $\chi^2$).

For $\delta_{12}(\sqrt{s})$, a satisfactory description is obtained to $\sim$
1.52 GeV with $p=0.948201+0.31767i$ (this corresponds to the pole on the
$s$-plane at $s=0.434 {\rm GeV}^2$). Here 59 experimental points \cite{Wickl}
are considered; $\chi^2/\mbox{ndf}\approx 3.05$ when eliminating the points
at 1.117, 1.247, and 1.27 GeV (with an especially large contribution to
$\chi^2$). The total $\chi^2/\mbox{ndf}$ for four analyzed quantities to
describe the processes $\pi\pi\to\pi\pi,K\overline{K}$ is 2.12; the number of
adjusted parameters is 17.
In Table~\ref{tab:clusters}, the obtained pole clusters for resonances are
shown on the complex energy plane ($\sqrt{s_r}={\rm E}_r-i\Gamma_r$).
\begin{table}[ht]
\centering
\caption{
Pole clusters for resonances obtained in solution 1.
}
\vskip0.3truecm
\begin{tabular}{|c|rl|rl|rl|rl|rl|rl|}
\hline
{} & \multicolumn{2}{c|}{$f_0 (665)$} & \multicolumn{2}{c|}{$f_0(980)$}
& \multicolumn{2}{c|}{$f_0(1500)$} \\
\cline{2-7}
Sheet & \multicolumn{1}{c}{E, MeV} & \multicolumn{1}{c|}{$\Gamma$, MeV}
& \multicolumn{1}{c}{E, MeV} & \multicolumn{1}{c|}{$\Gamma$, MeV}
& \multicolumn{1}{c}{E, MeV} & \multicolumn{1}{c|}{$\Gamma$, MeV}\\
\hline
II & 610$\pm$14 & 620$\pm$26 & 988$\pm$5 & 27$\pm$8 & 1530$\pm$25
& 390$\pm$30 \\
\hline
III & 720$\pm$15 & 55$\pm$9 & 984$\pm$16 & 210$\pm$22 & 1430$\pm$35
& 200$\pm$30 \\
{} & {} & {} & {} & {} & 1510$\pm$22~ & 400$\pm$34 \\
\hline
IV & {} & {} & {} & {} & 1410$\pm$24 & 210$\pm$38  \\
\hline \end{tabular}
\label{tab:clusters}
\end{table}

Now we calculate the coupling constants of these states with $\pi\pi$ and
$K\overline{K}$ systems ($g_1$ and $g_2$, respectively) through the residues
of amplitudes at the pole on sheet II, expressing the $T$-matrix via the
$S$-matrix as
$~S_{ii}=1+2i\rho_i T_{ii}$, $~ S_{12}=2i\sqrt{\rho_1\rho_2} T_{12}$,
where $\rho_i=\sqrt{(s-4m_i^2)/s}$, and taking the resonance part of
amplitude in the form
$~T_{ij}^{res}=\sum_r g_{ir}g_{rj}D_r^{-1}(s),$
where $D_r(s)$ is the inverse propagator ($D_r(s)\propto s-s_r$).
We obtain (in GeV units):
for $f_0(665)$: $g_1=0.7477\pm 0.095$ and $g_2=0.834\pm 0.1$, for
$f_0(980)$: $g_1=0.1615 \pm 0.03$ and $g_2=0.438 \pm 0.028$, for
$f_0(1500)$: $g_1=0.899 \pm 0.093$.

{\bf Solution 2} (analysis with ${f_0}(1370)$): The description of the
$\pi\pi$ scattering from the threshold to 1.89 GeV is practically the same as
without the $f_0(1370)$: $\chi^2/\mbox{ndf}$ for $\delta_1$ and $\eta$ is
2.01. The description of data is slightly improved for $|S_{12}|$
which is reached now up to $\sim$ 1.46 GeV. For this quantity, we now
consider 41 experimental points \cite{Wickl}; $\chi^2/\mbox{ndf}\approx 0.92$.
However, on the whole, the description is even worse than with the 1st
solution: the total $\chi^2/\mbox{ndf}\approx 2.93$ for four analyzed
quantities to describe the processes $\pi\pi\to\pi\pi,K\overline{K}$ (cf.
2.12 for the 1st case). The number of adjusted parameters is 21 where they
all  are positions of the poles describing resonances except a single one
related to the $K\overline{K}$ background which is $p=0.976745+0.214405i$
(this corresponds to the pole on the $s$-plane at $s=0.622 {\rm GeV}^2$).

In Table~\ref{tab:clusters2}, the obtained clusters for considered resonances
are shown (the cluster for the $f_0(665)$ remains the same as in
solution 1, therefore, it is not shown in Table~\ref{tab:clusters2}).
\begin{table}[ht]
\centering
\caption{
Pole clusters for resonances obtained in solution 2.
}
\vskip0.3truecm
\begin{tabular}{|c|rl|rl|rl|rl|rl|rl|}
\hline
{} & \multicolumn{2}{c|}{$f_0(980)$}
& \multicolumn{2}{c|}{$f_0(1370)$} & \multicolumn{2}{c|}{$f_0(1500)$} \\
\cline{2-7}
Sheet & \multicolumn{1}{c}{E, MeV} & \multicolumn{1}{c|}{$\Gamma$, MeV}
& \multicolumn{1}{c}{E, MeV} & \multicolumn{1}{c|}{$\Gamma$, MeV}
& \multicolumn{1}{c}{E, MeV} & \multicolumn{1}{c|}{$\Gamma$, MeV}\\
\hline
II & 986$\pm$5 & 25$\pm$8 & {} & {} & 1530$\pm$22
& 390$\pm$28 \\
\hline
III & 984$\pm$16 & 210$\pm$25 & 1340$\pm$21
& 380$\pm$25 & 1490$\pm$30 & 220$\pm$25 \\
& {} & {} & {} & {} & 1510$\pm$22~ & 370$\pm$30 \\
\hline
IV & {} & {} & 1330$\pm$18 & 270$\pm$20 & 1490$\pm$20 & 300$\pm$35\\
\hline \end{tabular}
\label{tab:clusters2}
\end{table}
When calculating the coupling constants in solution 2, we should take, for
the ${f_0}(1370)$, the residues of amplitudes at the pole on sheet IV.
We obtain (in GeV units):
for $f_0(665)$: $g_1=0.724\pm 0.095$ and $g_2=0.582\pm 0.1$, for
$f_0(980)$: $g_1=0.153 \pm 0.03$ and $g_2=0.521\pm 0.028$, for
$f_0(1370)$: $g_1=0.11\pm 0.03$ and $g_2=0.91\pm 0.04$, for
$f_0(1500)$: $g_1=0.866 \pm 0.09$.

So, the ${f_0}(980)$ and especially the ${f_0}(1370)$ are coupled essentially
more strongly to the $K\overline{K}$ system than to the $\pi\pi$ one. This
tells about the dominant $s{\bar s}$ component in the ${f_0}(980)$ state and
especially in the ${f_0}(1370)$ one.

Let us also indicate the scattering lengths calculated for both the
solutions. For the $K\overline{K}$ scattering:
$~a_0^0(K\overline{K})=-1.188\pm 0.13+(0.648\pm 0.09)i,~
[m_{\pi^+}^{-1}]; ~~~({\rm  solution 1}),$\\
$~~~~~~~~~~~~~~~~~~~~~~~~a_0^0(K\overline{K})=-1.497\pm 0.12+(0.639\pm 0.08)i,~
[m_{\pi^+}^{-1}]; ~~~({\rm solution 2}).$\\
The imaginary part in $a_0^0(K\overline{K})$ means that, already at the
threshold of the $K\overline{K}$ scattering, other
channels ($2\pi,4\pi$ etc.) are opened. We see that the real part of the
$K\overline{K}$ scattering length is very sensitive to the existence of
the ${f_0}(1370)$ state.

In Table~\ref{tab:pipi.length}, we compare our results for the $\pi\pi$
scattering length $a_0^0$, obtained for both solutions, with results of some
other works both theoretical and experimental.
\begin{table}
\centering
\caption{
Comparison of results of various works for the $\pi\pi$ scattering
length $a_0^0$.
}
\vskip0.3truecm
\begin{tabular}{|c|l|l|} \hline
$a_0^0, ~m_{\pi^+}^{-1}$ & ~~~~~~~References & ~~~~~~~~~~~~~~~~~Remarks \\
\hline
$0.27\pm 0.06$ (1)& our paper & model-independent approach \\
$0.266$~~~~~~~~(2)&{}&{}\\
\hline
$0.26\pm 0.05$ & L. Rosselet et al.\cite{Hyams} & analysis of the decay
$K\to\pi\pi e\nu$ \\
{} & {} & using Roy's model\\
\hline
$0.24\pm 0.09$ & A.A. Bel'kov et al.\cite{Hyams} & analysis of
$\pi^-p\to\pi^+\pi^-n$ \\
{} & {} & using the effective range formula\\
\hline
$0.23$ & S. Ishida et al.\cite{Ishida} & modified analysis of
$\pi\pi$ scattering \\
{} & {} & using Breit-Wigner forms \\
\hline
$0.16$ & S. Weinberg \cite{Weinberg} & current algebra (non-linear
$\sigma$-model) \\
\hline
$0.20$ & J. Gasser, H. Leutwyler \cite{Gasser} & one-loop corrections,
non-linear\\
{} & {} & realization of chiral symmetry \\
\hline
$0.217$ & J. Bijnens at al.\cite{Bijnens} & two-loop
corrections, non-linear\\
{} & {} & realization of chiral symmetry  \\
\hline
$0.26$ & M.K. Volkov \cite{Volkov} & linear realization of chiral symmetry
\\
\hline
$0.28$ & A.N. Ivanov, N.I. Troitskaya \cite{Ivanov} & a variant of chiral
theory with\\
{} & {} & linear realization of chiral symmetry
\\
\hline
\end{tabular}
\label{tab:pipi.length}
\end{table}
We see that our results correspond to the linear realization of chiral
symmetry.

Here, we presented model-independent results. Masses and widths of these
states that should be calculated from the obtained pole positions and
coupling constants are highly model-dependent. Let us demonstrate this.

If $f_0(665)$ is the $\sigma$-meson, then from the known relation~
$g_{\sigma\pi\pi}=(m_\sigma^2-m_\pi^2)/\sqrt{2}f_{\pi^0}$~
($f_{\pi^0}=93.1$ MeV), we obtain ~$m_\sigma\approx 342$ MeV.
If we take the resonance part of amplitude as~
$T^{res}=\sqrt{s}\Gamma/(m_\sigma^2-s-i\sqrt{s}\Gamma)$,
we obtain $m_\sigma\approx 850$ MeV and $\Gamma\approx 1240$ MeV.

\noindent{\bf 4. Summary.}
It is shown that the large $\pi\pi$-background usually obtained in various
analyses combines, in reality, the influence of the left-hand branch-point
and the contribution of a wide resonance at $\sim$ 665 MeV. Thus, a
model-independent confirmation of the state, denoted in the PDG issues by
$f_0(400-1200)$ \cite{PDG-00} is obtained.

A parameterless description of the $\pi\pi$ background is given by
allowance for the left-hand branch-point in the proper uniformizing variable.
Therefore, all the adjusted parameters in describing the $\pi\pi$ scattering
are the positions of poles corresponding to resonances,
and we conclude that our model-independent approach is a valuable tool for
studying the realization schemes of chiral symmetry.
The existence of $f_0(665)$ and the obtained $\pi\pi$-scattering length
($a_0^0(\pi\pi)\approx 0.27$) suggest the linear realization of chiral
symmetry.

The analysis of the used experimental data evidences that, if the
${f_0}(1370)$ resonance exists (soluiton 2), it has the dominant $s{\bar s}$
component, because the ratio of its coupling constant with the $\pi\pi$
channel to the one with the $K\overline{K}$ channel is 0.12.
A minimum scenario of the simultaneous description of processes
$\pi\pi\to\pi\pi,K\overline{K}$ does without the ${f_0}(1370)$ resonance.
The $K\overline{K}$ scattering length is very sensitive to whether this
state exists or not.

The $f_0 (1500)$ state is represented by the pole cluster which corresponds to
a flavour singlet, {\it e.g.} the glueball.

We think that multichannel states are most adequately represented by
clusters, {\it i.e.}, by the pole positions on all the corresponding sheets.
The pole positions are rather stable characteristics for various models,
whereas masses and widths are very model-dependent for wide resonances.


\begin{thebibliography}{99}

\bibitem{KMS-nc96} D. Krupa, V.A. Meshcheryakov, Yu.S. Surovtsev.
Nuovo Cim. {\bf A109}, 281 (1996).
\bibitem{PDG-00} Review of Particle Physics. Europ. Phys. J. {\bf C15}, 1
(2000).
\bibitem{NJL} Y. Nambu, G. Jona-Lasinio. Phys. Rev. {\bf 122}, 345 (1961);
M.K. Volkov. Ann. Phys. {\bf 157}, 282 (1984);
T. Hatsuda, T. Kunihiro. Phys. Rep. {\bf 247}, 223 (1994);
R. Delbourgo, M.D. Scadron. Mod. Phys. Lett. {\bf A10}, 251 (1995).
\bibitem{LN} K.J. Le Couteur. Proc. Roy. Soc. {\bf A256}, 115 (1960);
R.G. Newton. J. Math. Phys.{\bf 2}, 188 (1961).
\bibitem{Hyams} B. Hyams {\em et al.} Nucl. Phys. {\bf B64}, 134 (1973);
ibid. {\bf B100}, 205 (1975);
A. Zylbersztejn {\em et al.} Phys. Lett. {\bf B38}, 457 (1972);
P. Sonderegger, P. Bonamy {\sl Proc. 5th Intern. Conf. on Elementary
Particles} (Lund, 1969) paper 372;
J.R. Bensinger {\em et al.} Phys. Lett. {\bf B36}, 134 (1971);
J.P. Baton {\em et al.} Phys. Lett. {\bf B33}, 525, 528 (1970);
P. Baillon {\em et al.} Phys. Lett. {\bf B38}, 555 (1972);
L. Rosselet {\em et al.} Phys. Rev. {\bf D15}, 574 (1977);
A.A. Kartamyshev {\em et al.} Pis'ma v Zh. Eksp. Teor. Fiz. {\bf 25}, 68
(1977);
A.A. Bel'kov {\em et al.} Pis'ma v Zh. Eksp. Teor. Fiz. {\bf 29}, 652 (1979).
\bibitem{Wickl} A.B. Wicklund {\em et al.} Phys. Rev. Lett. {\bf 45}, 1469
(1980);
D. Cohen {\em et al.} Phys. Rev. {\bf D22}, 2595 (1980);
A. Etkin {\em et al.} Phys. Rev. {\bf D25}, 1786 (1982).
\bibitem{skn-app} Yu.S. Surovtsev, D. Krupa, M.Nagy. Acta Physica Polonica
{\bf B31}, 2697 (2000).
\bibitem{Ishida} S. Ishida {\em et al}. Progr. Theor. Phys. {\bf 95}, 745
(1996); ibid. {\bf 98}, 621 (1997).
\bibitem{Weinberg} S. Weinberg. Phys. Rev. Lett. {\bf 17}, 616 (1966);
B.W. Lee, H.T. Nieh. Phys. Rev. {\bf 166}, 1507 (1968).
\bibitem{Gasser} J. Gasser, H. Leutwyler. Ann. Phys. {\bf 158}, 142 (1984).
\bibitem{Bijnens} J. Bijnens {\em et al.}. Phys. Lett. {\bf B374}, 210 (1996).
\bibitem{Volkov} M.K. Volkov. Phys. Elem. Part. Atom. Nuclei {\bf 17, part 3},
433 (1986).
\bibitem{Ivanov} A.N. Ivanov, N.I. Troitskaya. Nuovo Cim. {\bf A108}, 555
(1995).

\end{thebibliography}
\end{document}